\documentclass[12pt]{article}

\begin{document}
\pagenumbering{arabic}
\title{The electromagnetic field in accelerated frames}

\author{J. W. Maluf$\,^{(a)}$ and  F. F. Faria$\,^{(b)}$ \\
Instituto de F\'{\i}sica, \\ 
Universidade de Bras\'{\i}lia, \\
C.P. 04385 \\
70.904-970 Bras\'{\i}lia, DF, Brazil\\}

\date{}
\maketitle
\begin{abstract}
We develop a geometrical framework that allows to obtain the electromagnetic
field quantities in accelerated frames. The frame of arbitrary accelerated
observers in space-time is defined by a suitable set of tetrad fields, whose
timelike components are adapted to the worldlines of a field of observers.
We consider the Faraday tensor and Maxwell's equations as abstract tensor
quantities in space-time, and make use of tetrad fields to project the
electromagnetic field quantities in the accelerated frames. As an application,
plane and spherical electromagnetic waves are projected in linearly accelerated
frames in Minkowski space-time. We show that the amplitude, frequency and the 
wave vector of the plane wave in the accelerated frame vary with time, while 
the light speed remains constant. We also obtain the variation of the Poynting
vector with time in the accelerated frame.
\end{abstract}
\bigskip
\noindent (a) wadih@unb.br, jwmaluf@gmail.com\par
\noindent (b) fff@unb.br\par
\bigskip
\section{Introduction}
Maxwell's electromagnetic theory is more than a century old. It is a well
established and understood theory. Usually the theory is presented in 
standard textbooks as a field theory in flat space-time 
\cite{Landau,Jackson}. The establishment of the theory in curved space-time 
\cite{Hehl2} requires the understanding of how exactly the
Faraday tensor couples with the gravitational field, and presently this is an
open issue. In the ordinary description of electrodynamics in flat space-time 
one almost always assumes that the sources and fields are established in an 
inertial reference frame. Very few investigations \cite{Mashhoon1} attempt to 
extend the analysis to accelerated frames. Such extension is mandatory because 
most frames in nature are, in one or another way, accelerated.

Until recently the attempts to describe the electromagnetic field in an 
accelerated frame consisted in performing a coordinate transformation of the
Faraday tensor defined in an inertial frame in flat space-time. For this 
purpose one considers a coordinate transformation from the flat space-time 
cartesian
coordinates to coordinates that describe a hyperbola in Rindler space, in case
of uniform acceleration. This procedure is not satisfactory for two reasons. 
First, a coordinate transformations is not a frame transformation. A coordinate
transformation is carried out on vectors and tensors on a manifold, and they 
just express the fact that (i) a point on the manifold may be labelled by 
different coordinates in different charts, and that (ii) one can work with any 
set of coordinates. On the other hand, a frame transformation is a Lorentz 
tranformation, it satisfies the properties of the Lorentz group and is carried 
out in the tangent space of the manifold. 

The second reason is that by considering an accelerated frame as a frame
obtained by a coordinate transformation, one cannot provide satisfactory 
answers to situations that are eventually understood as paradoxes, because the
inertial and ``accelerated" fields are described in different coordinates. One
of these paradoxes is the following: are the two situations, (i) an accelerated
charge in an inertial frame, and (ii) a charge at rest in an inertial frame 
described from the perspective of an accelerated frame, physically equivalent?

The procedure to be considered here consists, first, in assuming that the 
Faraday tensor and Maxwell's equations are abstract tensor quantities in 
space-time. Then we make use of tetrad fields to project the electromagnetic 
field either on an inertial or on a non-inertial frame, in the same coordinate 
system, in flat space-time. Tetrad fields constitute a set of four orthonormal 
vectors, that are adapted to observers that follow arbitrary paths in 
space-time. They constitute the local frame of these observers. Since the 
fields in the inertial frame and in the accelerated frame are defined in the 
same coordinate system, they can be compared with each other  unambiguously.

Given any set of tetrad fields, we may construct the acceleration tensor, as we
will show. This tensor determines the inertial (i.e., non-gravitational) 
accelerations that act on a given observer. For instance, a stationary observer
in space-time undergoes inertial forces, otherwise it would follow a geodesic 
motion determined by the gravitational field. A given frame (or a given tetrad 
field) may be characterized by the inertial accelerations.

In this chapter we will obtain the general form of Maxwell's equations that 
hold in inertial or noninertial frames. The formalism ensures that the 
procedure for projecting electromagnetic fields in noninertial frames is 
mathematically and physically consistent, and allows the investigation of 
several paradoxes. It is possible to conclude, for instance, that the 
radiation of an accelerated charged particle in an inertial frame is different
from the radiation of the same charged particle measured in a frame that is
co-accelerated (equally accelerated) with the particle.
Consequently, the accelerated motion in space-time is not 
relative, and the radiation of an accelerated charged particle is an absolute 
feature of the theory \cite{Maluf1}. 

We will study in detail the description of plane and spherical electromagnetic
waves in linearly accelerated frames in Minkowski space-time. We will show that
(i) the amplitude, (ii) the frequency and the wave vector of the plane wave, 
and (iii) the Poynting vector in the accelerated frame vary (decrease) with 
time, while the light speed remains constant.

\bigskip
\noindent Notation:

\begin{enumerate}
\item
Space-time indices $\mu, \nu, ...$ and Lorentz (SO(3,1)) indices
$a, b, ...$ run from 0 to 3. Time and space indices are indicated according to
$\mu=0,i,\;\;a=(0),(i)$.
\item
The space-time is flat, and therefore the metric tensor in 
cartesian coordinates is given by $g_{\mu\nu}= (-1, +1, +1, +1)$
\item
The tetrad field is 
represented by $e^a\,_\mu$. The flat, tangent space Minkowski 
space-time metric tensor raises and lowers tetrad indices and is fixed
by $\eta_{ab}= e_{a\mu} e_{b\nu}g^{\mu\nu}=(-1,+1,+1,+1)$.
\item
The frame components are given by the inverse tetrads $e_a\,^\mu$,
although we may as well refer to $\lbrace e^a\,_\mu\rbrace$ as the frame. The 
determinant of the tetrad field is represented by $e=\det(e^a\,_\mu)$
\end{enumerate}

\section{Reference frames in space-time}

The electromagnetic field is described by the Faraday tensor $F^{\mu\nu}$. In
the present analysis we will consider that $\lbrace F^{\mu\nu}\rbrace$ are just
{\it tensor components} in the flat Minkowski space-time described by 
{\it arbitrary} coordinates $x^\mu$. The projection of $F^{\mu\nu}$ on inertial
 or noninertial frames yields the electric and magnetic fields $E_x$, $E_y$, 
$E_z$, $B_x$, $B_y$ and $B_z$, which are the {\it frame components} of 
$\lbrace F^{\mu\nu}\rbrace$.
The projection is carried out with the help of tetrad fields 
$e^a\,_\mu$. For instance, $E_x=-cF^{(0)(1)}$, where $c$ is the speed of light 
and $F^{(0)(1)}=e^{(0)}\,_\mu e^{(1)}\,_\nu F^{\mu\nu}$. The study of the
kinematical properties of tetrad fields is mandatory for the 
characterization of reference frames.

Tetrad fields constitute a set of four orthonormal vectors in space-time,
$\lbrace e^{(0)}\,_\mu, e^{(1)}\,_\mu, e^{(2)}\,_\mu, e^{(3)}\,_\mu\rbrace$, 
that establish the local reference frame of an observer that moves along a 
trajectory $C$, represented by functions $x^\mu(s)$ 
\cite{Hehl1,Maluf2,Maluf3} ($s$ is the proper time of the observer). The tetrad
field yields the space-time metric tensor $g_{\mu\nu}$ by means of the relation
$e^a\,_\mu e^b\,_\nu \eta_{ab}=g_{\mu\nu}$, and $e^{(0)}\,_\mu$ and 
$e^{(i)}\,_\mu$ are timelike and spacelike vectors, respectively. 

We identify the $a=(0)$ component of $e_{a}\,^{\mu}$ with the observer's 
velocity $u^{\mu}$ along the trajectory $C$, i.e., $e_{(0)}\,^{\mu}=u^{\mu}/c
= dx^{\mu}/(cd\tau)$. 
The observer's acceleration $a^{\mu}$ is given by the absolute 
derivative of $u^{\mu}$ along $C$,
\begin{equation}
a^{\mu} = \frac{D u^{\mu}}{d\tau} = c \, \frac{De_{(0)}\,\!^{\mu}}{d\tau}.
\label{1}
\end{equation}
The absolute derivative is constructed with the help of the Christoffel 
symbols. Thus $e_{(0)}\,^{\mu}$ and its absolute derivative determine the 
velocity and acceleration along the worldline of an observer adapted to the 
frame. The set of tetrad fields for which $e_{(0)}\,^{\mu}$ describes a 
congruence of timelike curves is adapted to a class of observers 
characterized by the velocity field $u^{\mu}=c\,e_{(0)}\,^{\mu}$ and by the 
acceleration $a^\mu$. If $e^{a}\,_{\mu} = \delta^{a}_{\mu}$ everywhere in 
space-time, then $e^{a}\,_{\mu}$ is adapted to static observers, and 
$a^{\mu} = 0$.

We may consider not only the acceleration of observers along trajectories 
whose tangent vectors are given by $e_{(0)}\,^{\mu}$, but the acceleration 
of the whole frame along $C$. The acceleration of the frame is determined 
by the absolute derivative of $e_{a}\,^{\mu}$ along the path $x^\mu(\tau)$. 
Thus, assuming that the observer carries a frame, the acceleration of the 
latter along the path is given by  \cite{Mashhoon1,Mashhoon4}, 
\begin{equation}
\frac{De_{a}\,^{\mu}}{d\tau} = \phi_{a}\,^{b}e_{b}\,^{\mu},
\label{2}
\end{equation}
where $\phi_{ab}$ is the antisymmetric acceleration tensor of the 
frame ($\phi_{ab} = -\phi_{ba}$). According to 
Refs. \cite{Mashhoon1,Mashhoon4}, 
in analogy with the Faraday tensor we can identify 
$\phi_{ab} \equiv (\vec{a}/c,\vec{\Omega})$, where $\vec{a}$ is 
the translational acceleration ($\phi_{(0)(i)}=a_{(i)}/c$) and 
$\vec{\Omega}$ is the frequency of rotation ($\phi_{(i)(j)}= 
\epsilon_{(i)(j)(k)}\Omega^{(k)}$) of the spatial frame with respect 
to a nonrotating (Fermi-Walker transported \cite{Hehl1,Maluf3}) frame. 
It follows from Eq. (2) that
\begin{equation}
\phi_{a}\,^{b} = e^{b}\,_{\mu} \frac{D e_{a}\,^{\mu}}{d\tau}.
\label{3}
\end{equation}
Therefore given any set of tetrad fields for an arbitrary space-time, 
its geometrical interpretation may be obtained by suitably interpreting 
the velocity field $u^\mu=e_{(0)}\,\!^{\mu}$ and the acceleration tensor 
$\phi_{ab}$.

Using the definiton of the absolute derivative, we can write Eq. (3) as
\begin{eqnarray}
\phi_{a}\,^{b} &=& e^{b}\,_{\mu}\left( \frac{d e_{a}\,^{\mu}}{d\tau} 
+ \Gamma^{\mu}\,_{\lambda \sigma} \frac{d x^{\lambda}}{d \tau} 
e_{a}\,^{\sigma} \right) = e^{b}\,_{\mu} \left( 
\frac{dx^{\lambda}}{d\tau}\frac{\partial e_{a}\,^{\mu}}{\partial 
x^{\lambda}}+ \Gamma^{\mu}\,_{\lambda \sigma} 
\frac{d x^{\lambda}}{d \tau} e_{a}\,^{\sigma} \right) \nonumber \\ 
&=& e^{b}\,_{\mu} u^{\lambda} \left(\frac{\partial e_{a}\,^{\mu}}
{\partial x^{\lambda}}+ \Gamma^{\mu}\,_{\lambda \sigma}
e_{a}\,^{\sigma} \right) =  e^{b}\,_{\mu}u^{\lambda}\nabla_{\lambda} 
e_{a}\,^{\mu}.
\label{4}
\end{eqnarray}
Following Ref. \cite{Maluf2}, we take into account the orthogonality 
of the tetrads and write Eq. (4) as $\phi_{a}\,^{b}= - u^\lambda 
e_{a}\,^{\mu} \nabla_\lambda e^{b}\,_{\mu}$, where $\nabla_\lambda 
e^{b}\,_{\mu}=\partial_\lambda e^{b}\,_{\mu}-
\Gamma^{\sigma}\,_{\lambda \mu} e^{b}\,_{\sigma}$. Next we consider 
the identity $\partial_{\lambda}e^{b}\,_{\mu}-
\Gamma^{\sigma}\,_{\lambda \mu} e^{b}\,_{\sigma}+\,
^0\omega_{\lambda}\,^{b}\,\!_{c} \, e^{c}\,_{\mu}=0$, where 
$^0\omega_{\lambda}\,^{b}\,_{c}$ is the Levi-Civita spin connection given by
Eq. (21) below, and express $\phi_{a}\,^{b}$ according to
\begin{equation}
\phi_{a}\,^{b}= u^{\lambda} e_{a}\,^{\mu} 
\left(\,^0\omega_{\lambda}\,^{b}\,_{c} \, e^{c}\,_{\mu} \right) = c 
\,e_{(0)}\,^{\mu}(\,^0\omega_{\mu}\,^{b}\,_{a}).
\label{5}
\end{equation}
Finally we make use of  the identity $^0\omega_{\mu}\,^{a}\,_{b}= - 
K_{\mu}\,^{a}\,_{b}$, where $K_{\mu}\,^{a}\,_{b}$ is the contortion 
tensor defined by
\begin{equation}
K_{\mu ab}=\frac{1}{2}e_{a}\,^{\lambda} e_{b}\,^{\nu}(T_{\lambda \mu\nu}+
T_{\nu\lambda\mu}+T_{\mu\lambda\nu}),
\label{6}
\end{equation}
where
\begin{equation}
T^{\lambda}\,_{\mu\nu}=e_{a}\,^{\lambda} T^{a}\,_{\mu\nu} = 
e_{a}\,^{\lambda}\left( \partial_{\mu}e^{a}\,_{\nu} - \partial_{\nu}
e^{a}\,_{\mu}  \right),
\label{7}
\end{equation} 
is the object of anholonomity. Note that $T^{\lambda}\,_{\mu\nu}$ 
is also the torsion tensor of the Weitzenb\"ock space-time. After simple 
manipulations we arrive at
\begin{equation}
\phi_{ab} = \frac{c}{2} \left[ T_{(0)ab}+T_{a(0)b}-T_{b(0)a} \right],
\label{8}
\end{equation}
where $T_{abc} = e_{b}\,^{\mu}e_{c}\,^{\nu}T_{a\mu\nu}$.
The expression above is not covariant under local Lorentz (SO($3,1$) 
or frame) transformations, but is invariant under coordinate 
transformations. The noncovariance under local Lorentz transformations 
allows us to take the values of $\phi_{ab}$ to characterize the frame.

In order to measure field quantities with magnitude and direction (velocity, 
acceleration, etc.), an observer must project these quantities on the frame 
carried by the observer. The projection of a vector $V^{\mu}$ on a particular
frame is determined by
\begin{equation}
V^{a}(x) = e^{a}\,_{\mu}(x)\,V^{\mu}(x),
\label{9}
\end{equation}
and the projection of a tensor $T^{\mu\nu}$ is
\begin{equation}
T^{ab}(x) = e^{a}\,_{\mu}(x)\,e^{b}\,_{\nu}(x)\,T^{\mu\nu}(x).
\label{10}
\end{equation}
Note that the projections are carried out in the same coordinate system. 

We consider now an accelerated observer that follows a worldline 
$\bar{x}^{\mu}(\tau)$ in Minkowski space-time and carries a tetrad 
$e^{a}\,_{\mu}$, such that $e_{(0)}\,^{\mu} = u^{\mu}/c$ and 
$De_{a}\,^{\mu}/d\tau = \phi_{a}\,^{b}e_{b}\,^{\mu}$. At each instant 
$\tau$ of proper time along the worldline there are spacelike geodesics 
orthogonal to the worldline that form a local spacelike hypersurface. 
The observer can assign local coordinates 
$x^{a} =\lbrace x^{(0)},x^{(i)}\rbrace= \{c\tau, \vec{x'} \}$ 
to an event, which is also described by Cartesian 
coordinates $x^{\mu} = \{ct, \vec{x} \}$ 
belonging to this hypersurface, where 
\begin{equation}  
x^{(0)} = c\tau, \ \ \ \ \ \ \ x^{(i)} = [x^{\mu} - \bar{x}^{\mu}]
e^{(i)}\,_{\mu}.
\label{11}
\end{equation}
The inverse transformation reads
\begin{equation}
x^{\mu} = \bar{x}^{\mu} + e_{(i)}\,^{\mu}x^{(i)}.
\label{12}
\end{equation}
If we differentiate both sides of this equation over the worldline, we find
\begin{eqnarray}
dx^{\mu} &=& \left( \frac{1}{c}\frac{d\bar{x}^{\mu}}{d \tau} + 
\frac{1}{c}\frac{d e_{(i)}\,^{\mu}}{d \tau} \, x^{(i)} \right) 
dx^{(0)}+ e_{(i)}\,^{\mu}dx^{(i)} \nonumber \\  &=& \left( 
e_{(0)}\,^{\mu} + \frac{1}{c}\phi_{(i)}\,^{a}\, e_{a}\,^{\mu}\,
x^{(i)} \right)  dx^{(0)} + e_{(i)}\,^{\mu}dx^{(i)}.
\label{13}
\end{eqnarray}
Substituting Eq. (13) into the line element $ds^{2} = \eta_{\mu\nu}dx^{\mu}
dx^{\nu}$, we obtain the metric in the local coordinate system of an 
accelerated observer,
\begin{eqnarray}
ds^{2} &=& \left[ - \left(1+ \frac{\vec{a} \cdot \vec{x'}}{c^2} \right)^{2}
+ \frac{1}{c^2}\left( \vec{\Omega} \times \vec{x'} \right)^2 \right]
(dx^{(0)})^{2} + \left(\frac{2}{c} \, \vec{\Omega} \times \vec{x'}\right) 
dx^{(0)}dx^{(i)} \nonumber \\ &&  + \, \eta_{(i)(j)}dx^{(i)}dx^{(j)},
\label{14}
\end{eqnarray}
where we used $\phi_{(i)}\,^{(0)} x^{(i)} = (\vec{a} \cdot \vec{x'})/c$ 
and $\phi_{(j)}\,^{(i)}x^{(j)} = \left( \vec{\Omega} \times 
\vec{x'} \right)^{(i)}$. 

We see from Eq. (14) that $\eta_{(0)(0)}\cong-1$ only in the regions of 
space-time where 
\begin{equation}
|\vec{x'}| \ll \frac{c^2}{|\vec{a}|}\;, \ \ \ \ \ \ \ \ 
\mbox{and} \ \ \ \ \ \ \ 
|\vec{x'}| \ll \frac{c}{|\vec{\Omega}|}\;\;.
\label{15}
\end{equation}
Furthermore, some $c\tau =\; \mbox{constant}$ surfaces will intersect each 
other if we extend the spatial local coordinates far away from the observer's
worldline, which is not an admissible situation. Since we cannot assign two 
sets of coordinates for the same event, the local spatial coordinates have a 
limit of validity. In fact, the local coordinate system of Eq. (11) is valid 
only in those regions in the neighborhood of the observer's wordline in which
Eqs. (15) hold. We call $c^2/|\vec{a}|$ the translational acceleration 
length and $c/|\vec{\Omega}|$ the rotational acceleration length. On the 
Earth's surface, for example, we have ($|\vec{a}| = 9,8 \, 
\mbox{m}/\mbox{s}^2$, $|\vec{\Omega}| = \Omega_{\oplus}$)
\begin{equation}
\frac{c^2}{|\vec{a}|} = 9.46\cdot10^{15} \, \mbox{m} \approx 1 \,\mbox{ly} 
\ \ \ \ \ \ \ \mbox{and} \ \ \ \ \ \ \ \frac{c}{|\vec{\Omega}|} = 4.125
\cdot 10^{12}\, \mbox{m} \approx 27.5 \, \mbox{AU}.
\label{16}
\end{equation}
Hence we can use the local coordinates $x^{a}$ with confidence in most 
experimental situations in a laboratory on the Earth, where $|\vec{x'}|$ is 
negligible comparing to the acceleration lengths.

\section{The formulation of Maxwell's theory in moving frames}

The vector potential $A^\mu$, the Faraday tensor 
$F_{\mu\nu}=\partial_\mu A_\nu - \partial_\nu A_\mu$ and the four-vector 
current $J^\mu$ are vector and tensor components in space-time. Space-time
indices are raised and lowered by means of the flat space-time metric tensor 
$g_{\mu\nu}=(-1,+1,+1,+1)$. On a particular frame the electromagnetic 
quantities are projected and measured according to 
$A^a(x)=e^a\,_\mu(x) A^\mu(x)$ and 
$F^{ab}(x)=e^a\,_\mu(x)e^a\,_\nu(x) F^{\mu\nu}(x)$.

An inertial frame is characterized by the vanishing of the acceleration tensor 
$\phi_{ab}$. A realization of an inertial frame in Minkowski space-time is 
given by $e^a\,_\mu(t,x,y,z)=\delta^a_\mu$. It is easy to verify that this 
frame satisfies $\phi_{ab}=0$. More generally, all tetrad fields that are 
function of space-time {\it independent} parameters (boost and rotation 
parameters) determine inertial frames. Suppose that $A^a$ are componentes of 
the vector potential in an inertial frame, i.e., 
$A^a=(e^a\,_\mu)_{in}A^\mu=\delta^a_\mu A^\mu$. The components of $A^a$ in a 
noninertial frame are obtained by means of a local Lorentz transformation, 

\begin{equation}
\tilde{A}^a(x) = \Lambda^a\,_b(x) A^b(x)\,,
\label{17}
\end{equation}
where $\Lambda^a\,_b(x)$ are space-time dependent matrices that satisfy

\begin{equation}
\Lambda^a\,_c(x) \Lambda^b\,_d(x)\eta_{ab}=\eta_{cd}\,.
\label{18}
\end{equation}
In terms of covariant indices we have $\tilde{A}_a(x)=\Lambda_a\,^b(x)A_b(x)$.
An alternative but completely equivalent way of obtaining the field components 
$\tilde{A}_a(x)$ consists in performing a frame transformation by means of a 
suitable noninertial frame $e^a\,_\mu$, namely, in projecting $A^\mu$ on the 
noninertial frame,

\begin{equation}
\tilde{A}^a(x)=e^a\,_\mu(x) A^\mu(x)\,.
\label{19}
\end{equation}
Of course we have 
$\Lambda^a\,_b \,\delta^b_\mu= \Lambda^a\,_b\,(e^b\,_\mu)_{in}= e^a\,_\mu$.

The covariant derivative of $A_a$ is defined by

\begin{eqnarray}
D_a A_b&=&e_a\,^\mu D_\mu A_b \nonumber \\
&=&e_a\,^\mu (\partial_\mu A_b-{\;}^0\omega_\mu\,^c\,_b A_c)\,, 
\label{20}
\end{eqnarray}
where 

\begin{eqnarray}
^0\omega_{\mu ab}&=&-{1\over 2}e^c\,_\mu(
\Omega_{abc}-\Omega_{bac}-\Omega_{cab})\,, \nonumber \\
\Omega_{abc}&=&e_{a\nu}(e_b\,^\mu\partial_\mu e_c\,^\nu-
                        e_c\,^\mu\partial_\mu e_b\,^\nu)\,,
\label{21}
\end{eqnarray}
is the metric-compatible Levi-Civita connection considered in Eq. (5).
Note that we are considering
the flat space-time, and yet this connection may be nonvanishing. In 
particular, for noninertial frames it is nonvanishing. The Weitzenb\"ock 
torsion tensor $T^a\,_{\mu\nu}$ is also nonvanishing. However, the
curvature tensor constructed out of ${}^0\omega_{\mu ab}$ vanishes
identically: $R^a\,_{b\mu\nu}({\,}^0\omega)\equiv 0$. 

Under a local Lorentz transformation the spin connection transforms as

\begin{equation}
\widetilde{^0\omega}_\mu\,^a\,_b=
\Lambda^a\,_c({\,}^0\omega_\mu\,^c\,_d)\Lambda_b\,^d
+\Lambda^a\,_c\partial_\mu \Lambda_b\,^c\,.
\label{22}
\end{equation}
It follows from eqs. (17), (21) and (22) that under a local Lorentz 
transformation we have

\begin{equation}
\tilde{D}_a\tilde{A}_b=\Lambda_a\,^c(x)\Lambda_b\,^d(x)\,D_cA_d\,.
\label{23}
\end{equation}

The natural definition of the Faraday tensor in a noninertial frame is 

\begin{equation}
F_{ab}=D_aA_b-D_bA_a\,.
\label{24}
\end{equation}
In view of eq. (24) we find that the tensors $F_{ab}$ and $\tilde{F}_{ab}$ 
in two arbitrary frames are related by

\begin{equation}
\tilde{F}_{ab}=\Lambda_a\,^c(x)\Lambda_b\,^d(x)F_{cd} \,.
\label{25}
\end{equation}

The Faraday tensor defined by eq. (24) is related to the standard expression
defined in inertial frames. By substituting (20) in (24) we find

\begin{eqnarray}
F_{ab}&=&e_a\,^\mu(\partial_\mu A_b-{\,}^0\omega_\mu\,^m\,_b A_m)-
e_b\,^\mu(\partial_\mu A_a-{\,}^0\omega_\mu\,^m\,_a A_m)  \nonumber \\
{}&=& e_a\,^\mu (\partial_\mu A_b)-e_b\,^\mu (\partial_\mu A_a)+
({\,}^0\omega_{abm}-{\,}^0\omega_{bam})A^m\,.
\label{26}
\end{eqnarray}
We make use of the {\it identity}

\begin{equation}
{\,}^0\omega_{abm}-{\,}^0\omega_{bam}=T_{mab}\,,
\label{27}
\end{equation}
where $T_{mab}$ is given by eq. (7), and write

\begin{eqnarray}
F_{ab}&=&e_a\,^\mu e_b\,^\nu(\partial_\mu A_\nu - \partial_\nu A_\mu)+
T^m\,_{ab}A_m \nonumber \\
&+&e_a\,^\mu(\partial_\mu e_b\,^\nu)A_\nu-
e_b\,^\mu(\partial_\mu e_a\,^\nu)A_\nu\,.
\label{28}
\end{eqnarray}
In view of the orthogonality of the tetrad fields we have

\begin{equation}
\partial_\mu e_b\,^\nu=-e_b\,^\lambda(\partial_\mu e^c\,_\lambda)e_c\,^\nu\,.
\label{29}
\end{equation}
With the help of the equation above we find that the last two terms of eq. 
(28) may be rewritten as 

\begin{equation}
e_a\,^\mu(\partial_\mu e_b\,^\nu)A_\nu-
e_b\,^\mu(\partial_\mu e_a\,^\nu)A_\nu = -T^m\,_{ab}A_m\,.
\label{30}
\end{equation}
Therefore the last three terms of (28) cancel each other and finally we have

\begin{equation}
F_{ab}=e_a\,^\mu e_b\,^\nu(\partial_\mu A_\nu - \partial_\nu A_\mu)\,.
\label{31}
\end{equation}

The equation above shows that given the abstract, tensorial expression of the
Faraday tensor we can simply project it on any moving frame in Minkowski
space-time. This is exactly the procedure adopted by Mashhoon \cite{Mashhoon1}
in the investigation of electrodynamics in accelerated frames. Mashhoon is 
interested in developing the non-local formulation of electrodynamics. 
However, if we restrict attention to the evaluation of total quantities, such
as the integration of the Poynting vector and the total radiated power
(and not to pointwise measurements), then the standard formulation suffices to
arrive at qualitative conclusions.

We may obtain Maxwell's equations with sources from an action integral 
determined by the Lagrangian density

\begin{equation}
L=-{1\over 4} e\,F^{ab}F_{ab}- \mu_0\,e\,A_b J^b \,,
\label{32}
\end{equation}
where $e=\det(e^a\,_\mu)$, $J^b=e^b\,_\mu J^\mu$ and $\mu_0$ is the magnetic
permeability constant. Although in flat space-time we have $e=1$, we keep $e$
in the expressions below because it allows a straightforward inclusion of 
the gravitational field. Note that in view of eq. (31) we have

\begin{equation}
F^{ab}F_{ab}=F^{\mu\nu}F_{\mu\nu}\,.
\label{33}
\end{equation}
Therefore $L$ is frame independent, besides being invariant under coordinate
transformations. The field equations derived from $L$ are

\begin{equation}
\partial_\mu(e\,F^{\mu b})+ e\,F^{\mu c}\,({}^0\omega_\mu\,^b\,_c) 
=\mu_0\, e\,J^b\,,
\label{34}
\end{equation}
or

\begin{equation}
e_b\,^\nu \lbrack
\partial_\mu(e\,F^{\mu b})+ e\,F^{\mu c}\,({}^0\omega_\mu\,^b\,_c) \rbrack
=\mu_0\, e\,J^\nu\,,
\label{35}
\end{equation}
where $F^{\mu c}=e_b\,^\mu F^{bc}$. In view of eq. (33) it is clear that the 
equations above are equivalent to the standard form of Maxwell's equations in
flat space-time.

The second set of Maxwell's equations is obtained by working out the quantity
$D_aF_{bc}+D_bF_{ca}+D_cF_{ab}$, where the covariant derivative of $D_aF_{bc}$
is defined by 

\begin{eqnarray}
D_a F_{bc}&=&e_a\,^\mu D_\mu F_{bc} \\ \nonumber
&=& e_a\,^\mu(\partial_\mu F_{bc}-{\,}^0\omega_\mu\,^m\,_b F_{mc}-
{\,}^0\omega_\mu\,^m\,_c F_{bm})\,.
\label{36}
\end{eqnarray}
Taking into account relations (27) and (29) we find that the source free 
Maxwell's equations in an arbitrary moving frame are given by

\begin{equation}
D_aF_{bc}+D_bF_{ca}+D_cF_{ab}= e_a\,^\mu e_b\,^\nu e_c\,^\lambda
(\partial_\mu F_{\nu\lambda}+\partial_\nu F_{\lambda \mu}+
\partial_\lambda F_{\mu\nu})=0\,,
\label{37}
\end{equation}
in agreement with the standard description.

We refer the reader to Ref. \cite{Maluf1}, where we consider an accelerated
frame with velocity $v(t)$ with respect to an inertial frame, and describe 
Gauss law in the accelerated frame for the situations (i) in which the 
source is at rest in the inertial frame, and (ii) in which the source is
at rest in the accelerated frame.

\section{Plane electromagnetic waves in a linearly accelerated frame}

In this section we consider an observer in Minkowski space-time that is 
uniformly accelerated in the positive $x$ direction. The wordline and 
velocity of the observer in terms of its proper time $\tau$ are
\begin{equation}
\bar{x}^{\mu} = \left\{ \frac{c^2}{a}\, \mathrm{sinh} \left(\frac{a\tau}{c}
\right), \frac{c^2}{a}\left[ \mathrm{cosh} \left(\frac{a\tau}{c} 
\right)-1 \right], 0, 0\right\},
\label{38}
\end{equation}
and
\begin{equation}
u^{\mu} = \frac{d\bar{x}^{\mu}}{d\tau} = \left\{ c \, \mathrm{cosh} 
\left(\frac{a\tau}{c} \right), c \, \mathrm{sinh} \left(\frac{a\tau}{c}
\right), 0, 0\right\},
\label{39}
\end{equation}
respectively. 

A simple form of tetrad fields adapted to the observer with velocity $u^\mu$,
i.e., for which $e_{(0)}\,^{\mu} = u^{\mu}/c$ and 
$e^{a}\,_{\mu}e_{a\nu} = \eta_{\mu\nu}$, is given by
\begin{equation}
e^{a}\,_{\mu}=\left(
\begin{array}{cccc}
  \mathrm{cosh}(a\tau/c) & - \, \mathrm{sinh}(a\tau/c) & 0 & 0 \\
  - \,  \mathrm{sinh}(a\tau/c)  & \mathrm{cosh}(a\tau/c) & 0 & 0 \\
  0 & 0 & 1 & 0 \\
  0 & 0 & 0 & 1 \\
\end{array}
\right).
\label{40}
\end{equation}
If we substitute the tetrad fields and the inverses into Eq. (4), we see 
that the only nonvanishing component of $\phi_{ab}$ is
\begin{equation}
\phi_{(0)}\,\!^{(1)} =  \frac{a}{c}.
\label{41}
\end{equation}
The frame described by Eq. (40) is moving with uniform acceleration $a$
in the positive $x$ direction, and its axes are oriented along the global
Cartesian frame. In view of Eqs. (12), (38) and (40), it follows that
\begin{eqnarray}
t &=& \frac{c}{a}\left(1 + \frac{ax'}{c^2}\right)\mathrm{sinh}\left(
\frac{a\tau}{c} \right) , \nonumber \\ x &=& \frac{c^2}{a}\left(  1+
\frac{ax'}{c^2}\right)\mathrm{cosh}\left(\frac{a\tau}{c} \right) - 
\frac{c^2}{a}, \nonumber \\
y&=& y', \nonumber \\
z&=& z'.
\label{42}
\end{eqnarray}
We note that Eq. (39) can be given alternatively in terms of the time
coordinate $t$ of the inertial frame by 

\begin{equation}
u^\mu(t) = \left\{ c\gamma(t),\,c\beta(t)\,\gamma(t), \,0,\, 0\right\}\,,
\label{43}
\end{equation}
where

$$
\gamma(t)=\sqrt{1+a^2t^2/c^2}\;, \ \ \ \ \ \ \ \ \mbox{and} \ \ \ \ \ \ \ 
\beta(t)\gamma(t)= at/c\;\;. $$

In terms of the coordinates $(t,x,y,z)$ adapted to the inertial frame, the
Faraday tensor for a plane electromagnetic wave that propagates in the
positive $x$ direction reads

\begin{equation}
F^{\mu\nu}=\left(
\begin{array}{cccc}
  0 & 0 & -E_{y}/c & 0 \\
  0  & 0 & -B_{z} & 0 \\
  E_{y}/c & B_{z} & 0 & 0 \\
  0 & 0 & 0 & 0 \\
\end{array}
\right).
\label{44}
\end{equation}
where 

\begin{equation}
E_{y}(t,\vec{x}) = E_{0}\cos{(kx-\omega t)},
\label{45}
\end{equation}
\begin{equation}
B_{z}(t,\vec{x}) = \frac{E_{0}}{c}\cos{(kx-\omega t)}.
\label{46}
\end{equation}
In these expressions $k$ is the wave number and $\omega$ is the frequency of 
the wave, which are related by $k =|\vec{k}|=\omega/c$. The speed of 
propagation of the electromagnetic wave is 
\begin{equation}
v_{p} = \frac{\omega}{|\vec{k}|} = c\;.
\label{47}
\end{equation} 
The expression of the electromagnetic field in the inertial frame is formally
obtained out of Eqs. (45) and (46) by means of the tetrad field
$e^a\,_\mu=\delta^a_\mu$. However we will 
consider that (45) and (46) do represent the fields in the inertial frame.

In view of the expressions above we see that the only nonzero component 
of the Poynting vector
\begin{equation}
\vec{S} = \frac{1}{\mu_{0}}\vec{E} \times \vec{B},
\label{48}
\end{equation}
is given by
\begin{equation}
S_{x} = \frac{(E_{0})^{2}}{\mu_{0}c}\cos^{2}{(kx-\omega t)},
\label{49}
\end{equation}
where $\mu_{0}$ is the magnetic permeability constant. Thus the energy 
flux of the electromagnetic wave points in the same direction of the 
wave propagation.

In order to obtain the electric and magnetic field components of the
electromagnetic wave in the uniformly accelerated frame, we insert 
Eqs. (44) and (40) into $F^{ab}=e^a\,_\mu e^b\,_\nu F^{\mu\nu}$.
We obtain 
\begin{equation}
E_{(y)} = \mathrm{cosh}\left( \frac{a\tau}{c}\right)E_{y} - c 
\, \mathrm{sinh}\left( \frac{a\tau}{c}\right) B_{z},
\label{50}
\end{equation}
\begin{equation}
B_{(z)} = - \frac{1}{c} \, \mathrm{sinh}\left( \frac{a\tau}{c}\right)
E_{y} + \mathrm{cosh}\left( \frac{a\tau}{c}\right) B_{z},
\label{51}
\end{equation}
where $E_y$ and $B_z$ are given by (45) and (46), respectively.

In Eqs. (50) and (51) $E_y$ and $B_z$ are expressed in terms of the 
coordinates $(t,\,x)$. In order to 
present the electric and magnetic fields in terms of the coordinates
$(\tau, \vec{x'})$ of the accelerated frame we make use of Eq. (42).
We arrive at

\begin{equation}
E_{(y)}(\tau,\vec{x'}) = E_{0} \,e^{- a\tau/c}\cos{\left[k\left( 
e^{- a\tau/c}\right)x'- \frac{\omega c}{a}  
\left( 1 - e^{- a\tau/c} \right) \right]},
\label{52}
\end{equation}

\begin{equation}
B_{(z)}(\tau,\vec{x'}) = \frac{E_{0}}{c} \,e^{- a\tau/c}\cos{\left[k
\left(e^{- a\tau/c}\right)x'- \frac{\omega c}{a} 
\left(1 - e^{- a\tau/c} \right) \right]},
\label{53}
\end{equation}
where we used 
$$e^{- a\tau/c} = \mathrm{cosh}(a\tau/c) - \, 
\mathrm{sinh}(a\tau/c)\,.$$
The only nonzero component of the Poynting vector is

\begin{equation}
S_{(x)} = \frac{(E_{0})^{2}}{\mu_{0}c} \, e^{- 2a\tau/c} \cos^{2}{
\left[k\left( e^{- a\tau/c}\right)x'- \frac{\omega c}{a} 
\left( 1 - e^{- a\tau/c} \right) \right]},
\label{54}
\end{equation}
We see that the density of energy flux decreases in time by a factor 
$e^{- 2a\tau/c}$ in a frame that is uniformly accelerated in the same 
direction of the propagation of the electromagnetic wave.

The amplitudes in Eqs. (52) and (53) may be written as
\begin{equation}
E_{(0)} = E_{0} \,e^{- a\tau/c},
\label{55}
\end{equation}
\begin{equation}
B_{(0)} = \frac{E_{0}}{c} \,e^{- a\tau/c} = \frac{E_{(0)}}{c}.
\label{56}
\end{equation}
The identification of the wave number and of the frequency of the wave in
the accelerated frame is made by means of a projection of the wave vector 
$k_\mu=(-\omega/c,k,0,0)$ from the inertial to the accelerated frame,
according to $k^{a} =e^{a}\,_{\mu}k^{\mu}$. We recall that this
procedure is equivalent to performing a local Lorentz transformation where
the coefficients $\Lambda^a\,_b$ of the transformation satisfy
$e^a\,_\mu =\Lambda^a\,_b\,(e^b\,_\mu)_{in}=\Lambda^a\,_b\,\delta^b_\mu$.
Thus we have

\begin{equation}
k' = k\,e^{- a\tau/c},
\label{57}
\end{equation}
\begin{equation}
\omega' = \omega\,e^{- a\tau/c}.
\label{58}
\end{equation}

We conclude that the amplitude, wave number and frequency of the 
electromagnetic wave decrease in proper time by a factor $e^{-a\tau/c}$
in a frame that is uniformly accelerated in the same direction of the wave
propagation. We note that the observer will never reach the speed of light. 
Considering Eqs. (57) and (58) we see that the speed of propagation of the
electromagnetic wave in the uniformly accelerated frame is
\begin{equation}
v'_{p} = {{\omega'}\over{k'}}=
\frac{\omega\,e^{- a\tau/c}}{k\,e^{- a\tau/c}} = c\;.
\label{59}
\end{equation}
Therefore the speed of the electromagnetic wave is independent of the
observer's acceleration.

\section{Spherical waves in a radially accelerated frame}

We will repeat the analysis carried out in the previous section and consider
the measurement of spherical electromagnetic waves, produced in an inertial
frame, in a radially accelerated frame. A spherical wave in an inertial frame
may be characterized by the following expressions for the electric and magnetic
fields,

\begin{equation}
\vec{E}(t,r,\theta,\phi) = E_{0}\frac{\sin{\theta}}{r}\left[ 
\cos{(kr-\omega t)} - \frac{1}{kr}\sin{(kr-\omega t)} \right] 
{\hat{\phi}},
\label{60}
\end{equation}

\begin{equation}
\vec{B}(t,r,\theta,\phi) = - \frac{E_{0}}{c}\frac{\sin{\theta}}{r}\left[ 
\cos{(kr-\omega t)} - \frac{1}{kr}\sin{(kr-\omega t)} \right] 
{\hat{\theta}},
\label{61}
\end{equation}
where the unit vectors $\hat{\phi}$ and $\hat{\theta}$ are defined in terms of
the cartesian unit vectors by

\begin{eqnarray}
\hat{ \phi} &=& -\sin{\phi} \, \hat{\bf x} + \cos{\phi} 
\, \hat{\bf y}, \nonumber \\ 
\hat{\theta} &=& \cos{\theta}\cos{\phi} \, \hat{\bf x} + 
\cos{\theta}\sin{\phi} \, \hat{\bf y} - \sin{\theta} \,
\hat{\bf z}.
\label{62}
\end{eqnarray}

A set of tetrad fields in spherical coordinates, adapted to an observer that 
undergoes uniform acceleration in the radial direction, is given by

\begin{equation}
e^{a}\,_{\mu}(t,r,\theta,\phi) =\left(
\begin{array}{cccc}
  \gamma &  - \gamma\beta & 0 & 0 \\
   - \gamma\beta  & \gamma & 0 & 0 \\
  0 & 0 & r & 0 \\
  0 & 0 & 0 & r\sin{\theta} \\
\end{array}
\right),
\label{63}
\end{equation}
where

\begin{equation}
\gamma = \sqrt{1+\frac{a^{2}t^{2}}{c^{2}}}, \ \ \ \ \ \ \ \gamma\beta = 
\frac{at}{c}.
\label{64}
\end{equation}
The inverse components of Eq. (63) are such that 
$e_{(0)}\,^\mu(t,r,\theta,\phi)=(\gamma,\beta \gamma\,0,0)$.
Therefore the frame is accelerated along the radial direction.

We start with the Faraday tensor in cartesian coordinates,

\begin{equation}
F^{\mu\nu}(t,x,y,z)=\left(
\begin{array}{cccc}
  0 & -E_{x}/c & -E_{y}/c & -E_{z}/c \\
  E_{x}/c  & 0 & -B_{z} & B_{y} \\
  E_{y}/c & B_{z} & 0 & - B_{x} \\
  E_{z}/c & -B_{y} & B_{x} & 0 \\
\end{array}
\right).
\label{65}
\end{equation} 
The electric and magnetic field components in the expression above are 
obtained out of Eqs. (60), (61) and (62). We must consider the expression
above in spherical coordinates. So we perform the coordinate transformation

\begin{equation}
F'^{\alpha\beta}(t,r,\theta,\phi) = \frac{\partial x'^{\alpha}}{\partial 
x^{\mu}}\frac{\partial x'^{\beta}}{\partial x^{\nu}}F^{\mu\nu}(t,x,y,z).
\label{66}
\end{equation}
After some algebra we obtain

\begin{eqnarray}
F'^{01} &=& 0, \nonumber \\
F'^{02} &=& 0, \nonumber \\ 
F'^{03} &=& - \frac{E_{0}}{c}\frac{1}{r^2}\left[ \cos{(kr-\omega t)} 
- \frac{1}{kr}\sin{(kr-\omega t)} \right], \nonumber \\
F'^{12} &=& 0, \nonumber \\
F'^{23} &=& 0, \nonumber \\
F'^{31} &=&  \frac{E_{0}}{c}\frac{1}{r^2}\left[ \cos{(kr-\omega t)} 
- \frac{1}{kr}\sin{(kr-\omega t)} \right].
\label{67}
\end{eqnarray}
The quantities in Eq. (67) represent both the abstract tensor
components of the Faraday tensor in spherical coordinates, and the components
of the Faraday tensor in an inertial frame. Next we project these tensor
components on the accelerated frame defined by Eq. (63). We arrive at

\begin{eqnarray}
F'^{(0)(1)} &=& 0, \nonumber \\
F'^{(0)(2)} &=& 0, \nonumber \\ 
F'^{(0)(3)} &=& - \frac{E_{0}}{c}(\gamma - \gamma\beta)\frac{\sin{\theta}}{r}
\left[ \cos{(kr-\omega t)} - \frac{1}{kr}\sin{(kr-\omega t)} \right], 
\nonumber \\
F'^{(1)(2)} &=& 0, \nonumber \\
F'^{(2)(3)} &=& 0, \nonumber \\
F'^{(3)(1)} &=& \frac{E_{0}}{c}(\gamma - \gamma\beta)\frac{\sin{\theta}}{r}
\left[ \cos{(kr-\omega t)} - \frac{1}{kr}\sin{(kr-\omega t)} \right].
\label{68}
\end{eqnarray}
Note that the factor $(\gamma -\gamma\beta)$ may be rewritten as

\begin{equation}
\gamma -\gamma\beta = \sqrt{ {{1-\beta}\over{1+\beta}}}\,.
\label{69}
\end{equation}

In order to verify how Eqs. (60) and (61) are modified in the accelerated
frame we just compare the structure of Eqs. (67) and (68), and indentify 
(60) and (61) in the latter expression. We obtain

\begin{equation}
\vec{E}(t,r,\theta,\phi) = E_{0} (\gamma-\beta\gamma) 
{{\sin{\theta}}\over r}
\left[ \cos{(kr-\omega t)} - \frac{1}{kr}\sin{(kr-\omega t)} \right] 
{\hat{\phi}},
\label{70}
\end{equation}

\begin{equation}
\vec{B}(t,r,\theta,\phi) = - \frac{E_{0}}{c}(\gamma - \beta\gamma)
{{\sin{\theta}}\over r}
\left[ \cos{(kr-\omega t)} - \frac{1}{kr}\sin{(kr-\omega t)} \right] 
{\hat{\theta}},
\label{71}
\end{equation}
in the inertial frame coordinates.

By comparing Eqs. (70) and (71) with (60) and (61) we see that the major
qualitative difference between these expressions is the emergence, in the 
former pair of equations, of the time dependent Doppler factor 
$(\gamma - \beta\gamma)$ given by Eq. (69). If the accelerated frame is at
the radial position $(r, \theta)$ at the instant $t$, then the measured
amplitude of the wave in the accelerated frame will be smaller by a factor
$(\gamma -\beta\gamma)$ than if the frame were at rest at the same position.
Thus the amplitude of the
spherical wave in the accelerated frame varies with time, and approaches
zero in the limit $t\rightarrow \infty$, since in this limit 
$\beta \rightarrow 1$.

\section{Final comments}

The tetrad field and its interpretation as a frame adapted to arbitrary
observers in space-time allow the formulation of electrodynamics in
accelerated frames. The idea is to project the electromagnetic vectorial and
tensorial quantities in any moving frame by means of the tetrad field.
Specific issues regarding electromagnetic radiation were discussed in ref. 
\cite{Maluf1}. 

Of course all the results derived from the procedure adopted in this chapter 
are valid as long as the very concept of tetrad field and its 
interpretation are also valid. The justification behind the usage of tetrad
fields for this purpose is given by principle of locality \cite{Mashhoon5}.
The idea is the following. A physical measurement is considered to be
reliable if it is performed in an inertial reference frame. Normally it is 
admitted that the observer is standing in an inertial frame. 
Measurements in accelerated frame are, in general, not easily performed. 
When an electromagnetic field quantity is projected in a frame by means of the
tetrad field, it is assumed that this tetrad field is, at each instant of time,
physically equivalent (identical) to another frame that is inertial and 
momentarily co-moving with the accelerated frame. The worldline of the two
frames, the accelerated and the inertial, coincide at that instant of time.
To a certain extent, the hypothesis of locality, together with the concept of 
tetrad field, extends the principle of relativity, since it relates inertial 
and non-inertial frames.

An interesting consequence of the present analysis is the following. Let us 
suppose that an accelerated observer in the context of section 4 measures the
frequencies $\omega^\prime_1$ and $\omega^\prime_2$ at the instants of proper
time $\tau_1$ and $\tau_2$, respectively,

\begin{equation}
\omega'_{1} = \omega\,e^{- a\tau_{1}/c}, \ \ \ \ \ \ \ \omega'_{2} = 
\omega\,e^{- a\tau_{2}/c},
\label{72}
\end{equation}
according to eq. (58). By dividing the two frequencies of the electromagnetic
waves we obtain

\begin{equation}
a = \frac{c}{\Delta \tau} \ln{\left(\frac{\omega'_{1}}{\omega'_{2}}
\right)},
\label{73}
\end{equation} 
where $\Delta \tau = \tau_{2} - \tau_{1}$.
Therefore the accelerated observer may determine the value of its own
acceleration provided the luminosity of the source is constant and the 
acceleration is uniform. This formula may be useful in the evaluation of the
acceleration of the solar system, for instance, with respect to the distant
supernovas, provided it is verified that in the interval $\Delta \tau$ the
luminosity of the supernova is not substantially changed. Of course the
resulting value will provide just the order of magnitude of the acceleration
of the expansion of the universe.

The final expressions of the electric and magnetic fields in the accelerated
frames, Eqs. (52-53), and (70-71), for plane and spherical waves,
respectively, are related to the expressions in the inertial frame by means of
simple time dependent functions. The simplicity of the final expressions 
ensures that the present tehcnique is correct, and suggests that all
manifestations of electrodynamics may be investigated in any moving frame.

\end{document}